\documentclass[
    ,final            
  ]
  {aipproc}

\layoutstyle{6x9}

\begin{document}

\title{The Interaction Rate in Holographic Models of Dark Energy}



\author{Diego Pav\'{o}n}{
address={Departamento de F\'{\i}sica, Universidad Aut\'{o}noma de
Barcelona, 08193 Bellaterra, Spain}}

\author{Anjan A. Sen}{
address={Center for Theoretical Physics, Jamia Millia Islamia, New
Delhi 110025, India}}

\begin{abstract}
Observational data from supernovae type Ia, baryon acoustic
oscillations, gas mass fraction in galaxy clusters, and the growth
factor are used to reconstruct the the interaction rate of the
holographic dark energy model recently proposed by Zimdahl and
Pav\'{o}n \cite{wd} in the redshift interval $0 < z < 1.8$. It
shows a reasonable behavior as it increases with expansion from a
small or vanishing value in the far past and begins decreasing at
recent times. This suggests that the equation of state parameter
of dark energy does not cross the phantom divide line.
\end{abstract}

\maketitle

\section{Introduction}
There is a wide-shared conviction among cosmologists that the
Universe is nowadays experiencing a stage of accelerated expansion
not compatible with the up to now favored Einstein-de Sitter
model. This consensus, however, does not extend to the agent (very
frequently  called dark energy) behind this acceleration and, at
the moment, there are many competing candidates \cite{reviews},
the cosmological constant being the leading one. However, the
latter suffers from two main short-comings: the huge value
predicted by quantum field theory estimations, and the coincidence
problem, i.e., the fact that the energy densities of
non-relativistic matter and dark energy are currently of the same
order. This is why many researches are looking for alternative
candidates of dark energy. Among the most recent generic proposals
there is a very suggestive one based on the holographic principle.
Loosely speaking, the latter asserts that the entropy of a system
is given by the number of degrees of freedom lying on the surface
that bounds it, rather than in its volume \cite{principle}. The
roots of this principle are to be found in the thermodynamics of
black holes \cite{pedro-jakob}. Nevertheless, as observed by Cohen
{\em et al.} \cite{cohen}, a system may fulfill the holographic
principle and, however, include states for which its Schwarzschild
radius is larger than its size, $L$. This can be avoided by
imposing  that the energy of the system should not exceed that of
black hole of the same size or, equivalently, $\rho \leq 3 \,
c^{2}/(8\pi \, G \, L^{2})$, where $c^2$ is a (non-necessarily
constant) parameter. In the cosmological context $L$, the infrared
cutoff, is usually taken either as the event horizon radius or the
Hubble radius. For a quick summary of holographic dark energy see
section 3 of Ref. \cite{wd}.

The model of Zimdahl {\it et al.} \cite{wd} rests in two main
assumptions: $(i)$ dark energy complies with the holographic
principle with $L$ identified as the radius of the Hubble horizon,
$H^{-1}$, hence $\rho_{x} = 3c^{2} H^{2}/(8 \pi G)$, and $(ii)$
dark energy and dark matter do not evolve separately but they
interact with one another. As a consequence, the energy
conservation equations read
\begin{equation}
\dot{\rho}_{m} + 3H \rho_{m} = Q \, , \qquad \dot{\rho}_{x} + 3H
(1+w) \rho_{x} = - Q \, , \label{consv}
\end{equation}
\\
where $w = p_{x}/\rho_{x}$ is the equation of state parameter of
dark energy which is not constrained to be a constant. Subscripts
$m$ and $x$ are for non-relativistic dark matter and dark energy,
respectively.

It is to be noted that for spatially flat universes in the absence
of interaction, $Q = 0$, there would be no acceleration
\cite{wd,dwplb}. Moreover, $Q$ must be a positive-definite
quantity if the coincidence problem is to be solved \cite{rapid}
or at least alleviated \cite{alleviated}, and the second law of
thermodynamics to be fulfilled \cite{db}. Further if $Q$ were
negative, $\rho_{x}$ would have been negative in the far past.
Besides, it has been forcefully argued that the Layzer-Irvine
equation \cite{layzer} when applied to galaxy clusters reveals the
existence of the interaction \cite{elcio}. To the best of our
knowledge, the interaction hypothesis was first introduced, much
earlier of the discovery of late acceleration, by Wetterich
\cite{wetterich} to reduce the theoretical huge value of the
cosmological constant, and  was first used in the holography
context by Horvat \cite{raul}. As we write, the body of literature
on the subject is steadily growing -see \cite{wd,korean} and
references therein. Most cosmological models implicitly assume
that matter and dark energy couple gravitationally only. However,
unless there exists an underlying symmetry that would set $Q$ to
zero (such a symmetry is still to be discovered) there is no {\em
a priori} reason to discard the interaction. Ultimately,
observation will tell us whether the interaction exists.

Following \cite{wd}, we will write the interaction as $Q = \Gamma
\rho_{x}$, where $\Gamma$ is an unknown, semi-positive definite,
function that gauges the rate at which energy is transferred from
dark energy to dark matter. Clearly, as long as the nature of both
dark ingredients of the cosmic substratum remain unknown, $\Gamma$
cannot be derived from first principles. The alternative is to
resort to observational data (in our case, supernovae type Ia (SN
Ia), baryon acoustic oscillations (BAO), gas mass fraction in
galaxy clusters and the growth factor) to roughly reconstruct it
-for details see \cite{plbad}.

\section{Reconstructing the rate}
From Eqs. (\ref{consv}) and the above expressions for $\rho_{x}$
and $Q$ the evolution equation
\\
\begin{equation}
\dot{r} = (1+r) \, \left[  3 H w \frac{r}{1+r} + \Gamma \right] \,
, \label{revol1}
\end{equation}
\\
for the ratio $r \equiv \rho_{m}/\rho_{x}$ between the energy
densities, is readily obtained.  With the help of Friedmann
equation $\Omega_{m} + \Omega_{x} + \Omega_{k} = 1$, in terms of
the usual density parameters $\Omega_{i} = 8\pi G
\rho_{i}/(3H^{2}) \;$ ($i = m, x$), and $\Omega_{k} = - k/(a^{2}\,
H^{2})$, where $k$ stands for the spatial curvature index of the
Friedmann-Robertson-Walker metric, we can write
\\
\begin{equation}
\dot{r} = - 2 H\, \frac{\Omega_{k}}{\Omega_{x}} \, q \, \, ,
\label{revol2}
\end{equation}
\\
where $q = - \ddot{a}/(a\, H^{2})$ denotes the cosmic deceleration
parameter. Here, for the holographic dark energy we have adopted
the expression $\rho_{x}\propto H^2$. The latter follows from
choosing the infrared cutoff, $L$, as the Hubble radius, $H^{-1}$.

Likewise, starting from the first of Eqs. (\ref{consv}) and using
Friedmann equation, we get for the equation of state parameter
\\
\begin{equation}
w(z) = (1+r) \left[ \frac{2}{3} \frac{H'}{H}-1 \right]-
\frac{2}{3} \frac{\Omega_{k}}{\Omega_{x}}\left[ 1 -
(1+z)\frac{H'}{H} \right] \, , \label{wz1}
\end{equation}
\\
where $z$ denotes the redshift  and a prime indicates derivative
with respect to this quantity.

We fit the Chevallier-Polarsky-Linder parametrization \cite{w(z)},
namely,
\\
\begin{equation}
w(z)= w_{0} + w_{1} \frac{z}{1+z} \, , \label{wz2}
\end{equation}
\\
where $w_{0}$ is the present value of $w(z)$, and $w_{1}$ a
further constant, to current data from different observational
probes and subsequently use the fitting values for $w_{0}$ and
$w_{1}$ to reconstruct
 the dimensionless ratio $\Gamma/3H$.

As for the data, we resort to the various SN Ia observations in
recent times. In particular we use 60 Essence supernovae
\cite{super}, 57 SNLS (Supernova Legacy Survey) and  45 nearby
supernovae. We have also included the new data release of 30 SNe
Ia spotted by the Hubble Space Telescope and classified as the
Gold sample by Riess {\it et al.} \cite{super}. The combined data
set can be found in Ref. \cite{davis}. The total number of data
points involved is 192.

Next we add the measurement of the cosmic microwave background
(CMB) acoustic scale at $z_{BAO} = 0.35$ as observed by the Sloan
digital sky survey (SDSS)  for the large scale structure. This is
the  BAO peak \cite{sdss}.

We also consider the gas mass fraction of galaxy cluster, $f_{gas}
= M_{gas}/M_{tot}$, inferred from the x-ray observations
\cite{xrs}. This depends on the angular diameter distance $d_{A}$
to the cluster as $f_{gas} = d_{A}^{3/2}$. The number of data
points involved is 26.

Likewise, the two-degree field  galaxy redshift survey (2dFGRS)
has measured the two point correlation function at an effective
redshift of $z_{s} = 0.15$. This correlation function is affected
by systematic differences between redshift space and real space
measurements due to the peculiar velocities of galaxies. Such
distortions are expressed through the redshift distortion
parameter, $\beta$. Correlation function can be used to measure it
as $\beta = 0.49 \pm 0.09$ at the effective redshift of z = 0.15
of the 2dF survey. This result can be combined with linear bias
parameter $b= 1.04\pm 0.11$ obtained from the skewness induced in
the bispectrum of the 2dFGRS by linear biasing to find the growth
factor $g$ at $z = 0.15$, namely $g  = 0.51 \pm 0.11$
\cite{growth}.

\subsection{The spatially flat case}
\begin{figure}
\includegraphics[height=.6\textheight]{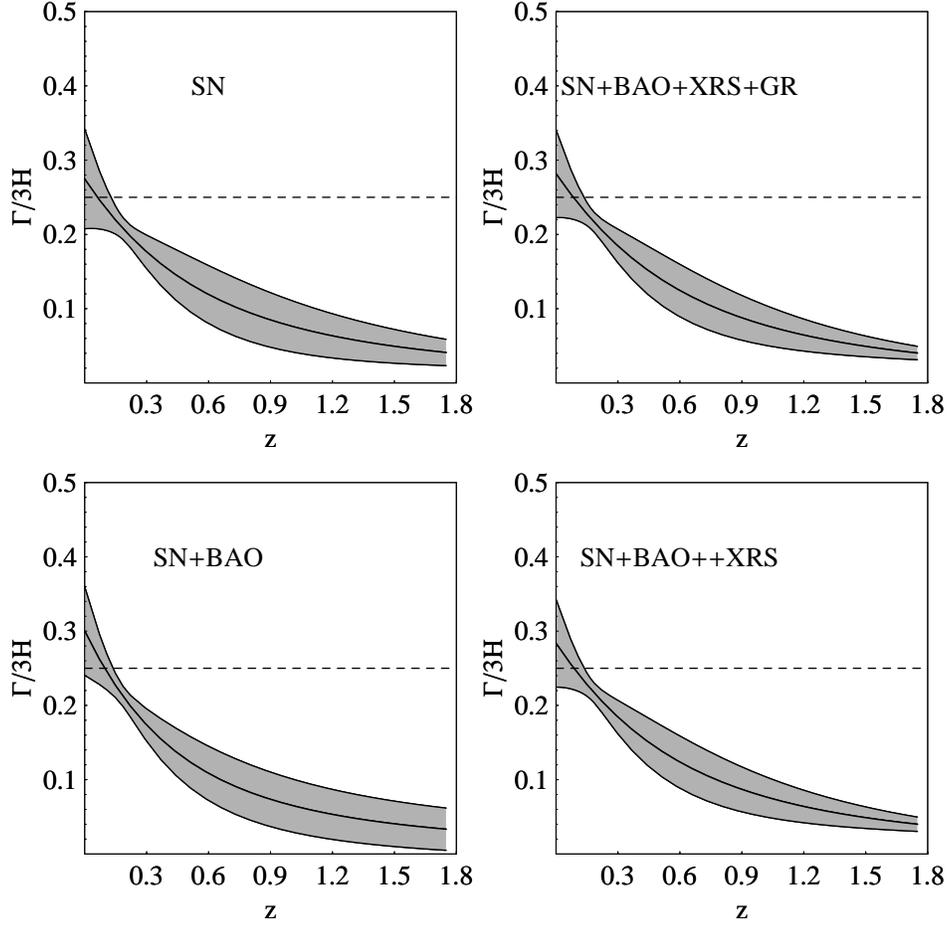}
\caption{The dimensionless ratio $\Gamma/(3H)$ vs redshift. In the
four panels we have fixed $\Omega_{m0} = 0.25$ and $\Omega_{k} =
0$. The solid line is for the mean value and the shaded area
indicates the $1\sigma$ region. The region above the horizontal
dashed line can be visited only when the dark energy becomes of
phantom type, i.e., $w < -1$. } \label{fig1}
\end{figure}

\begin{figure}
\includegraphics[height=.6\textheight]{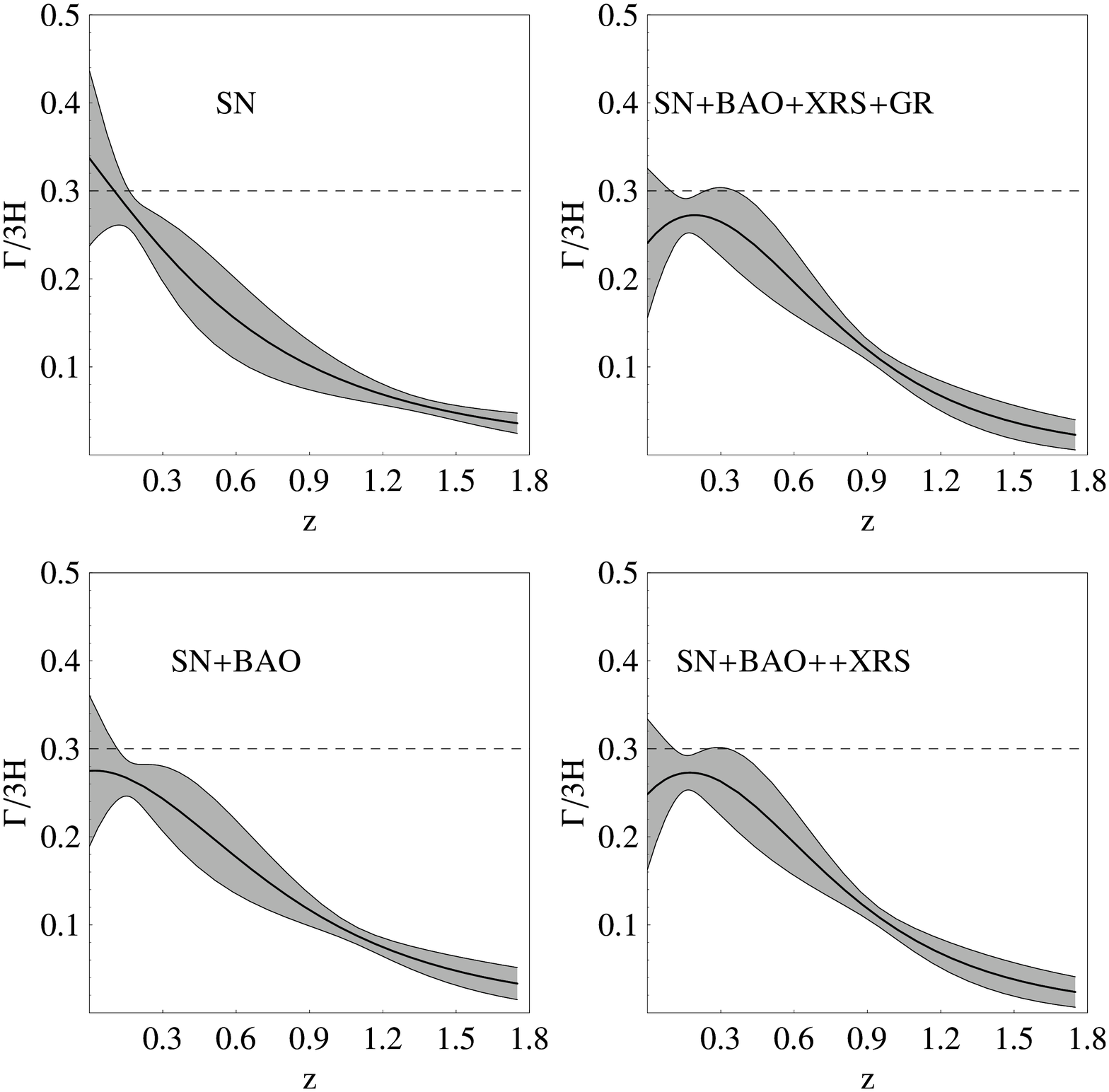}
\caption{Same as Fig. 1 except that here we have fixed
$\Omega_{m0} = 0.30$ and $\Omega_{k} = 0$.} \label{fig2}
\end{figure}

By setting $\Omega_{k} = 0$  equations (\ref{revol1}) and
(\ref{wz1}) reduce to
\\
\begin{equation}
\frac{\Gamma}{3H} = - r_{0} \left[  \frac{2}{3} \frac{H'}{H}- 1
\right] \, , \label{gammah1}
\end{equation}
\\
and
\\
\begin{equation}
w(z) = (1+ r_{0}) \left[  \frac{2}{3}  \frac{H'}{H}- 1\right]\, ,
\label{wz3}
\end{equation}
respectively\footnote{The corresponding equations (namely (6) \&
(7)) in \cite{plbad} bear an extra factor, $(1+z)$. This was a
typo with no further consequences to the calculations. This
corrects the typo.}. As usual, $r_{0}$ indicates the present value
of the ratio $r$. Using these two expressions we determine $w_{0}$
and $w_{1}$ from the data and, with them, we reconstruct $\Gamma/3
H$ -see figures \ref{fig1} and \ref{fig2}.
The best fit values, with $1\sigma$ error bars for the parameters
when all the data (SN Ia + BAO + x-rays + growth factor) are
included, come to be: \noindent $w_{0} = -1.13 \pm 0.24$, $w_{1} =
0.66 \pm 1.35$ (for $\Omega_{m0} = 0.25$ \& $\Omega_{k} = 0$, Fig.
\ref{fig1}); and  $w_{0} = -0.80 \pm 0.28$, $w_{1} = -1.75 \pm
1.79$ (for $\Omega_{m0} = 0.3$ \& $\Omega_{k} = 0$, Fig.
\ref{fig2}).

Contrary to what one may think, the fact that $r$ was never large
does not bring the model of Ref. \cite{wd} into conflict with the
standard scenario of cosmic structure formation. One may believe
that at early times the amount of dark matter would have been too
short to produce gravitational potential wells deep enough for
galaxies to condensate. However, this is not so; a matter
dominated phase is naturally recovered since at high and moderate
redshifts the interaction is even smaller than at present whence
the equation of state of the dark energy becomes close to that of
non-relativistic matter -see \cite{wd} for details.

Likewise, it should be noted that dark energy clusters similarly
to dark matter when the equation of state of the former stays
close to that of the latter. To see this more clearly we recall
the perturbation dynamics of this model studied in \cite{wd} by
using the perturbed metric $\mbox{d}s^{2} = - \left(1 + 2
\psi\right)\mbox{d}t^2 + a^2\,\left(1-2\psi\right)\delta _{\alpha
\beta} \mbox{d}x^\alpha\mbox{d}x^\beta \,$   with $\psi$ the
scalar metric perturbation, and the Bardeen gauge-invariant
variable \cite{bardeen}
\begin{equation}
\zeta \equiv -\psi + \frac{1}{3}\frac{\hat{\rho}}{\rho + p} =
-\psi - H \, \frac{\hat{\rho}}{\dot\rho}\, .
 \label{defzeta}
 \end{equation}
 The latter represents curvature perturbations on hypersurfaces of
 constant energy density. For the simplest case $\Gamma = $
 constant, one follows
\begin{equation}
\zeta = \zeta_{i} - \frac{\Gamma}{3}\frac{\hat{r}}{r}\frac{1}{3
H_{i} \, r - \Gamma}\left[\left(\frac{a}{a_{i}}\right)^{3/2} -
1\right]
 \ ,
\label{zeta}
\end{equation}
\\
where the subscript $i$ signals some initial time and a hat
indicates perturbation of the corresponding quantity -see
\cite{wd} and \cite{plbad} for details. Accordingly, as far as the
the equation of state parameter of dark energy $w$ remains close
to that of dark matter, $w \simeq 0$, both components cluster
similarly.

\subsection{Non-spatially flat cases}
It is immediately seen that for $\Omega_{k} \neq 0$ the ratio $r$
between energy densities is not a constant. This introduces a
further unknown function in our fitting procedure.
Notwithstanding, one should not expect a large variation in $r$ in
the redshift range $(0,1.8)$. In our subsequent computation, we
Taylor expand  $r$ around its present value up to the first order
term. Accordingly we parameterize it as
\begin{equation}
r = r_{0} + r_{1} \, \frac{z}{1+z} \, . \label{r(z)}
\end{equation}
Here $r_{1}$ is a constant which can be related to the present
ratio  between $\Omega_{k}$ and $\Omega_{x}$ by
\begin{equation}
\frac{\Omega_{k0}}{\Omega_{x0}} = - \frac{r_{1}}{2} \left[1 -
\left(\frac{H'}{H}\right)_{z = 0} \right]\, .
\label{presentdensitiesratio}
\end{equation}

This can be used to fix the unknown constant $r_{1}$ for a given
$\Omega_{k0}$ and $\Omega_{x0}$.  Also
\\
\begin{equation}
\frac{\Gamma}{3H} = - \frac{1}{1+r} \, \left(r' \, \frac{1+z}{3}-
w\, r \right) \, ,
\label{gammah2}
\end{equation}
\\
where $w$ is given by Eq. (\ref{wz1}).

With the help of these expressions, the ratio $\Gamma/3H$ can be
reconstructed from the data. The corresponding figure (with
$\Omega_{m0} = 0.30$ and small $\Omega_{k0}$) is very similar to
Fig. 2 thereby we do not reproduce it here (see Fig. 3 in Ref.
\cite{plbad}). The best fit values, with $1\sigma$ error bars for
the parameters when all the data (SN Ia + BAO + x-rays + growth
factor) are included, come to be: $w_{0} = -0.806 \pm 0.29$,
$w_{1} = -1.74 \pm 3.33$. We conclude that a small spatial
curvature, in agreement with the  WMAP 5yr experiment
 \cite{wmap5}, has a negligible impact on the evolution of
the interaction rate.

\section{Discussion}
Using the observational data (SNIa, BAO, gas mass fraction, and
growth factor) available in the redshift range $0< z< 1.8$ we
reconstructed the interaction term $Q$ of Ref. \cite{wd}. The
interaction rate $\Gamma$ (and hence $Q$) stays positive in the
said range. Its general trend is to decrease as $z$ increases but
it shows no indication of becoming negative at larger redshifts.
This corroborates that, as previously suggested \cite{db,elcio},
the energy transfer proceeds from dark energy to dark matter, not
the other way around. Although phantom behavior cannot be excluded
at recent and present times it only occurs  manifestly  either for
large $\Omega_{x0}$ -see Fig. \ref{fig1}- or when just the
supernovae data are considered (top-left panel of Figs. \ref{fig1}
and \ref{fig2}). When $\Omega_{x0}$ is somewhat lower (say, $0.7$)
and BAO and other data are included, the mean value of
dimensionless interaction rate, $\Gamma/3H$, no longer crosses the
phantom divide (i.e., the horizontal dashed line). It reaches a
maximum near $z = 0$ to subsequently decrease with expansion. This
result is rather comforting since holography does not seem
compatible with phantom energy \cite{bak}. On the other hand, it
should be noticed that $\Omega_{x0}$ values as high as $0.75$ do
not appear favored from a combination of results from WMAP 1yr and
weak lensing which yields $\Omega_{x0} = 0.70 \pm 0.3$
\cite{contaldi}.

Likewise, a small curvature term -of either sign- is of little
consequence.

At any rate, it is fair to say that the concordance $\Lambda$CDM
model ($w_{0} = -1$, $w_{1} = 0$) shows compatibility within
$1\sigma$ confidence level with the set of data considered in this
paper.

\begin{theacknowledgments}
We are grateful to the organizers of the Spanish Relativity
Meeting 2008, held in Salamanca (Spain),  for this opportunity.
This research was partly supported by the Spanish Ministry of
Science and Technology under Grant No. FIS2006-12296-C02-01, and
through a grant UAB-CIRIT, VIS-2007 for visiting professors.
\end{theacknowledgments}


\begin{thebibliography}{99}
\bibitem{wd}
W. Zimdahl and D. Pav\'{o}n, {\em Class. Quantum Grav.}
\textbf{24}, 5461 (2007).
\bibitem{reviews}
T. Padmanbahn, Phys. Rep. \textbf{380},  235 (2003);  V. Sahni,
Lect. Notes Phys. \textbf{653}, 141 (2004); J.A.S. Lima, {\em
Braz. J. Phys.} \textbf{34}, 194 (2004); L. Barnes, M.J. Francis,
G.F. Lewis, and E.V. Linder, {\em Publ. Astron. Soc. Australia}
\textbf{22}, 315 (2005); E.J. Copeland, M. Sami, and S. Tsujikawa,
{\em Int. J. Mod. Phys. D} \textbf{15}, 1753 (2006); L.
Perivolaropoulos, Third Aegean Summer School, {\em The Invisible
Universe Dark Matter and Dark Energy}, astro-ph/0601014; R. Durrer
and R. Maartens, {\em Gen. Relativ. Grav.} \textbf{40}, 301
(2008).
\bibitem{principle} G. `t Hooft, in {\em Dimensional Reduction in Quantum Gravity},
edited by A.Ali, J. Ellis, and S. Ranjbar-Daemi, Salmafestschrift:
A Collection of talks, World Scientific; Singapore, 1993,
gr-qc/9310026;  L. Susskind, {\em J. Math. Phys.} \textbf{36},
6377 (1995).
\bibitem{pedro-jakob}
J.D. Bekenstein, {\em Phys. Rev. D} \textbf{9}, 3292 (1974);\\
 {\em ibid.} \textbf{49}, 1912 (1994); \\
P. Gonz\'{a}lez-D\'{\i}az, {\em Phys. Rev. D} \textbf{27}, 3042
(1983).
\bibitem{cohen} A.G. Cohen, D.B. Kaplan, and A.E. Nelson, {\em Phys.
Rev. Lett.} \textbf{82}, 4971 (1999).
\bibitem{dwplb} D. Pav\'{o}n and W. Zimdahl, {\em Phys. Lett. B}
\textbf{628}, 206 (2005).
\bibitem{rapid} S. del Campo, R. Herrera and D. Pav\'{o}n, {\em Phys. Rev. D} \textbf{78}, 021302R (2008).
\bibitem{alleviated} W. Zimdahl, D. Pav\'{o}n, and L.P. Chimento,
{\em Phys. Lett. B} \textbf{521}, 133 (2001); L.P. Chimento, A.S.
Jakubi, D. Pav\'{o}n, and W. Zimdahl, {\em Phys. Rev. D}
\textbf{67}, 083513 (2003);  G. Olivares, F. Atrio-Barandela, and
D. Pav\'{o}n, {\em Phys. Rev. D} \textbf{71}, 063523 (2005); L.P.
Chimento and D. Pav\'{o}n, {\em Phys. Rev. D} \textbf{73}, 063511
(2006);  S. del Campo, R. Herrera and D. Pav\'{o}n, {\em Phys.
Rev. D} \textbf{74}, 023501 (2006).
\bibitem{db} J.S. Alca\~{n}iz and J.A.S. Lima, {\em Phys. Rev. D} \textbf{72}, 063516 (2005);
D. Pav\'{o}n and B. Wang, {\em Gen. Relativ. Grav.} (in the press)
arXiv:0712.0565 [gr-qc].
\bibitem{layzer} D. Layzer, {\em Astrophys. J.} \textbf{138}, 174
(1963); P.J.E. Peebles, {\em Principles of Cosmological Physics},
Princeton University Press, Princeton, New Jersey, 1993.
\bibitem{elcio} E. Abdalla,  L.R. Abramo, L.  Sodre, and
B. Wang,  arXiv:0710.1198 [astro-ph].
\bibitem{wetterich} C. Wetterich, {\em Nucl. Phys. B} \textbf{302}, 668
(1988).
\bibitem{raul} R. Horvat, {\em Phys. Rev. D} \textbf{70}, 087301 (2004).
\bibitem{korean} F. Atrio-Barandela and D. Pav\'{o}n,
``Interacting Dark Energy",  in {\em Dark Energy-Current Advances
and Ideas}, edit. J.R. Choi, Research Signpost, 2008 (in the
press).
\bibitem{plbad} A. A. Sen and D. Pav\'{o}n, {\em Phys. Lett. B}
\textbf{664}, 7 (2008).
\bibitem{w(z)} M. Chevallier and D. Polarski, {\em Int. J. Mod. Phys.
D} \textbf{10}, 213 (2001);  E.V. Linder, {\em Phys. Rev. Lett.}
\textbf{90}, 091301 (2003).
\bibitem{super} A.G. Riess {\em et al.}, {\em Astrophys. J.} {\bf 607}, 665
(2004); J.L. Tonry {\em et al.}, {\em Astrophys. J.} \textbf{594},
1 (2003); B.J. Barris {\em et al.}, {\em Astrophys. J.} {\bf 602},
571 (2004); P. Astier {\em et al.}, {\em Astron. Astrophys.} {\bf
447}, 31 (2006); A.G. Riess {\em et al.},  astro-ph/0611572.
\bibitem{davis} T. Davis {\em et al.}, astro-ph/0701510.
\bibitem{sdss} D.J. Eisenstein {\em et al.},  {\em Astrophys. J.} {\bf 633}, 560 (2005).
\bibitem{xrs} S.W. Allen, R.W. Schmidt, and A.C. Fabian,
{\em Mon. Not. R. Astron. Soc.} {\bf 334}, L11 (2002);  S.W. Allen
{\em et al.}, {\em Mon. Not. R. Astron. Soc.} \textbf{353}, 457
(2004).
\bibitem{growth}
L. Verde {\em et al.}, {\em Mon. Not. R. Astron. Soc.}
\textbf{335}, 432 (2002);  E. Hawkins {\em et al.}, {\em Mon. Not.
R. Astron. Soc.} \textbf{346}, 78 (2003).
\bibitem{bardeen}
J.M. Bardeen, P.J. Steinhardt, and M.S. Turner, {\em Phys. Rev. D}
\textbf{28}, 679 (1983).
\bibitem{wmap5} E. Komatsu {\it et al.}, arXiv:0803.0547 [astro-ph].
\bibitem{bak} D. Bak and S-J. Rey, {\em Class. Quantum Grav.}
\textbf{17}, L83 (2000);  E.E. Flanagan, D. Marolf, and R.M.Wald,
{\em Phys. Rev. D} \textbf{62}, 084305 (2000).
\bibitem{contaldi} C. Contaldi, H. Hoekstra, and A. Lewis, {\em Phys.
Rev. Lett.} \textbf{90}, 221303 (2003).
\end{thebibliography}
\end{document}